\documentclass[aps,pra,reprint,amsmath,amsfonts,amssymb,superscriptaddress]{revtex4-1}
\usepackage{graphicx,float,calc}
\usepackage{color,bm}
\usepackage{ulem}

\usepackage{braket}
\usepackage[colorlinks,urlcolor=blue,citecolor=blue,linkcolor=blue]{hyperref}
\begin{document}

\title{Critical properties in the non-Hermitian Aubry-Andr\'{e}-Stark model}

\author{Ji-Long Dong}
\affiliation{Key Laboratory of Atomic and Subatomic Structure and Quantum Control (Ministry of Education), Guangdong Basic Research Center of Excellence for Structure and Fundamental Interactions of Matter, South China Normal University, Guangzhou 510006, China}
\affiliation{Guangdong Provincial Key Laboratory of Quantum Engineering and Quantum Materials, School of Physics, South China Normal University, Guangzhou 510006, China}

\author{En-Wen Liang}
\affiliation{Key Laboratory of Atomic and Subatomic Structure and Quantum Control (Ministry of Education), Guangdong Basic Research Center of Excellence for Structure and Fundamental Interactions of Matter, South China Normal University, Guangzhou 510006, China}
\affiliation{Guangdong Provincial Key Laboratory of Quantum Engineering and Quantum Materials, School of Physics, South China Normal University, Guangzhou 510006, China}

\author{Shi-Yang Liu}
\affiliation{Key Laboratory of Atomic and Subatomic Structure and Quantum Control (Ministry of Education), Guangdong Basic Research Center of Excellence for Structure and Fundamental Interactions of Matter, South China Normal University, Guangzhou 510006, China}
\affiliation{Guangdong Provincial Key Laboratory of Quantum Engineering and Quantum Materials, School of Physics, South China Normal University, Guangzhou 510006, China}

\author{Guo-Qing Zhang}
\email{zhangptnoone@zjhu.edu.cn}
\affiliation{Research Center for Quantum Physics, Huzhou University, Huzhou 313000, People's Republic of China}

\author{Ling-Zhi Tang}
\email{tanglingzhi@quantumsc.cn}
\affiliation{Quantum Science Center of Guangdong-Hong Kong-Macao Greater Bay Area (Guangdong), Shenzhen 518045, China}

\author{Dan-Wei Zhang}
\email{danweizhang@m.scnu.edu.cn}
\affiliation{Key Laboratory of Atomic and Subatomic Structure and Quantum Control (Ministry of Education), Guangdong Basic Research Center of Excellence for Structure and Fundamental Interactions of Matter, South China Normal University, Guangzhou 510006, China}
\affiliation{Guangdong Provincial Key Laboratory of Quantum Engineering and Quantum Materials, School of Physics, South China Normal University, Guangzhou 510006, China}

\begin{abstract}
We explore the critical properties of the localization transition in the non-Hermitian Aubry-Andr\'{e}-Stark (AAS) model with quasiperiodic and Stark potentials, where the non-Hermiticity comes from the nonreciprocal hopping. The localization length, the inverse participation ratio and the energy gap are adopted as the characteristic quantities. We perform the scaling analysis to derive the scaling functions of the three quantities with critical exponents in several critical regions, with respect to the quasiperiodic and Stark potentials and the nonreciprocal strength. We numerically verify the finite-size scaling forms and extract the critical exponents in different situations. Two groups of new critical exponents for the non-Hermitian AAS model and its pure Stark limit are obtained, which are distinct to those for the non-Hermitian Aubry-Andr\'{e} model and their Hermitian counterparts. Our results indicate that the Hermitian and non-Hermitian AAS, Aubry-Andr\'{e}, and Stark models belong to different universality classes. We demonstrate that these critical exponents are independent of the nonreciprocal strength, and remain the same in different critical regions and boundary conditions. Furthermore, we establish a hybrid scaling function with a hybrid exponent in the overlap region between the critical regions for the non-Hermitian AAS and Stark models.

\end{abstract}

\date{\today}

\maketitle

\section{Introduction}

Localization transitions in quasiperiodic systems have attached broad interest in recent years \cite{Harper1955,Aubry1980,lellouch2014localization,devakul2017anderson, goblot_emergence_2020,agrawal_universality_2020,roy_critical_2022,agrawal_quasiperiodic_2022} . Compared to random quenched disorders, unconventional localization properties in systems with quasiperiodic disorders have been investigated both theoretically \cite{goblot_emergence_2020,agrawal_universality_2020,roy_critical_2022,agrawal_quasiperiodic_2022} and experimentally \cite{roati_anderson_2008,crespi_anderson_2013,bordia_coupling_2016}. Meanwhile, localization of wave functions can occur in clean systems without disorders, such as the Stark localization~\cite{Emin1987,Schulz2019,morong_observation_2021,chakrabarty2024fate} and flat-band localization~\cite{PhysRevB.95.115135,PhysRevLett.126.103601}. For disorder-induced localizations, the one-dimensional (1D) Aubry-Andr\'{e} (AA) model \cite{Aubry1980} with the quasiperiodic potential serves as an important toy model, whose critical point for the localization-delocalization transition can be determined by the self-duality method \cite{Biddle2010,liu_generalized_2020,wang_engineering_2023}. The original AA model has been intensively extended to the investigation of topological phases \cite{DWzhang2018,DWZhang2020,yoshii_topological_2021,GQZhang2021,tang_topological_2022,XLi2024,wang_topological_2022, nakajima_competition_2021,wu_quantized_2022,SHuang2024}, mobility edges \cite{Soukoulis1982,Ganeshan2015,TLiu2022,XCZhou2023}, many-body localization \cite{noauthor_observation_nodate,PhysRevLett.121.206601,lukin_probing_2019,wang_non-hermitian_2023,YWang2021}, and critical phenomena  \cite{agrawal_universality_2020,goblot_emergence_2020,lv_exploring_2022,lv_quantum_2022,aramthottil_finite-size_2021,devakul2017anderson,Roosz2024}. Continuous phase transition theory reveals universal phenomena near the critical point \cite{Osterloh2002,vojta2003quantum,Sachdev2011,Heyl2018,carollo_geometry_2020}. The universality is characterized by critical exponents near the critical point, which can be derived from scaling functions of physical observables in critical regions \cite{fisher_renormalization_1974,gosselin_renormalization_2001,belitz_how_2005,zhong_probing_2006,slevin_anderson_1997,asada_anderson_2002,lemarie_universality_2009,slevin_critical_2009,you_fidelity_2011,rams_quantum_2011, cherroret_how_2014,su_role_2018,wang_berezinskii-kosterlitz-thouless-like_2024}. Thus, the critical exponents in scaling functions classify phase transitions in different systems or models in terms of the universality class. For disordered systems, critical exponents play an important role in understanding localization transitions and corresponding critical phenomena. The universality class of localization transitions is usually determined by the spatial dimension and underlying symmetries of a system. Remarkably, new critical exponents for the localization transition in the disordered AA model with mixed random and quasiperiodic disorders have been revealed in Refs. \cite{S.Yin2022a,S.Yin2022b}. Different critical exponents for the localization transition in the Aubry-Andr\'{e}-Stark (AAS) model, which combines quasiperiodic potentials with Stark potentials, have also been uncovered \cite{ewliang2024,sahoo2024stark}.

In recent years, non-Hermitian systems have attracted increasing attention~\cite{Hatano1996,Hatano1997,Bender1998,Ueda2020,Kawabata2019,Zhou2019,Luo2021,Luo2022,Bergholtz2021}, where the non-Hermiticity comes from the gain-and-loss or nonreciprocality. Various intriguing non-Hermitian physics are discovered, such as the exceptional points ~\cite{Bergholtz2021,Shen2018,Xue2021,Tzeng2021,pnas.2302572120}, new types of topological states and invariants~\cite{Yao2018Aug,Yao2018Sep,Gong2018,Song2019,Jin2019,Xue2020,wang2021complex-energy,wang2021generating,DW2020TAI,Tang2020,Xue2022TAI}, and the non-Hermitian skin effect with skin modes localized near the boundaries under the open boundary condition (OBC)~\cite{Yao2018Aug,Longhi2019.Lett,Jiang2019,li2020critical,DW2020Skinsuperfluid,zhang2021observation, Zhang2021,YYi2020,LLi2020,zhang2022universal,Longhi2022,Longhi2022selfhealing,PhysRevB.106.014207,lin2023topological,SZLi2024NHSE} and the inner skin effect under the periodic boundary condition (PBC)~\cite{manna2023inner}. For localization in non-Hermitian systems, it has been found that the nonreciprocal hopping leads to delocalization \cite{Hatano1996,Hatano1997,Zhai2020,lztang2021} and the localization transitions coincide with topological and spectral
transitions \cite{Longhi2019.Lett,Jiang2019}. Several exotic localization properties have been predicted in the non-Hermitian AA model and quasicrystals, such as the generalized mobility edges \cite{Zeng2020,PhysRevB.109.054204,PhysRevA.110.012222,PhysRevB.101.174205,jiang2024exact}, the complex mobility rings \cite{LWang2024,SZLi2024}, the non-Hermitian quantum metric for revealing localization transition  points \cite{JFRen2024}, and the emergent entanglement phase transitions \cite{Kawabata2023,SZLi2024NHSE,LZhou2024}. The entanglement phase transition was also studied in the non-Hermitian Stark localization without disorders \cite{HZLi2024}. Recent studies have revealed that the non-Hermiticity enriches the tenfold Altland-Zirnbauer symmetry class into the 38-fold symmetry class \cite{Kawabata2019,Zhou2019}, which can lead to different critical exponents of localization transitions in the non-Hermitian and Hermitian disordered systems. For instance, new critical exponents have been obtained in 3D and 2D Anderson models with non-Hermitian random disorders \cite{Luo2021,Luo2022}, which demonstrate that the non-Hermiticity does change the universality class of the localization transitions. Thus, the symmetry classes enriched by the non-Hermiticity provide the physical origin of new critical exponents of non-Hermitian localization transitions. In addition, the hybrid scaling properties \cite{YMSun2024a} and the non-equilibrium dynamics \cite{YMSun2024b,S.Yin2022c} of the localization transition in the non-Hermitian disordered AA model were studied, and the critical exponents therein were shown to be distinct to the Hermitian counterpart \cite{S.Yin2022a,S.Yin2022b}. These works provide an interesting perspective on the quantum criticality of non-Hermitian localization transitions. However, it remains unexplored in other non-Hermitian systems with the combination of two different localization mechanisms.

\begin{figure}[t!]
	\centering
	\includegraphics[width=0.48\textwidth]{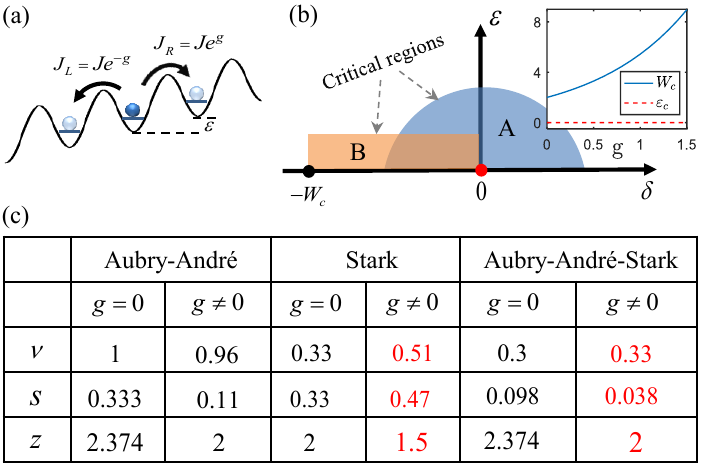}
	\caption{(Color online) (a) Illustration of the non-Hermitian AAS model with nonreciprocal hoppings and a linear gradient potential. (b) Sketch of the quantum criticality in the non-Hermitian AAS model. The critical region for (Stark and Anderson) localization transitions is denoted by blue region A. The orange region B stands for the critical region of the merely non-Hermitian Stark localization transition. These two critical regions overlap near the red dot labeled as the critical point at $W=W_{c}(g)$ and $\varepsilon=\varepsilon_{c}(g)$, which in the thermodynamic limit are plotted as functions of non-Hermitian parameter $g$. When $W=0$ for $\delta=W-W_c=-W_c$ labeled as the black dot, the model returns to the pure non-Hermitian Stark model. (c) Summary of the extracted critical exponents $\{\nu,s,z\}$ for the AA model, Stark model and AAS model under both Hermitian and non-Hermitian cases. The two groups of new critical exponents for $g\neq0$ are labeled in red.
	}\label{fig1}
\end{figure}

In this work, we explore the quantum criticality of the localization transition in the non-Hermitian AAS model with both quasiperiodic and Stark lattice potentials. Here the non-Hermiticity comes from the nonreciprocal hopping. We take the localization length, inverse participation ratio and energy gap as the physical quantities to characterize the critical behaviour of the localization transition. By extending the hybrid scaling method used in the Hermitian disordered AA and AAS models \cite{S.Yin2022a,S.Yin2022b,ewliang2024,sahoo2024stark}, we perform the scaling analysis to derive the scaling functions of the three quantities with critical exponents in several critical regions, with respect to the quasiperiodic and Stark potential strengths and the nonreciprocal strength. We then numerically verify these finite-size scaling functions and extract the critical exponents in different situations. As summarized in Fig. \ref{fig1}, our results indicate that the Hermitian and non-Hermitian AAS, AA, and Stark models with distinct critical exponents belong to different universality classes. Remarkably, we obtain two groups of new critical exponents for the non-Hermitian AAS model and its pure Stark limit, in contrast to those for the non-Hermitian AA model and their Hermitian counterparts. These critical exponents are independent of the nonreciprocal strength, and remain the same in different critical regions and boundary conditions. Furthermore, we establish a hybrid scaling function with a hybrid exponent in the overlap region between the critical regions for the non-Hermitian AAS and Stark models, where two scaling variables are relevant.

The rest of this paper is organized as follows. In Sec. \ref{sec2}, we introduce the non-Hermitian AAS model and the scaling analysis method. Section \ref{sec3} is devoted to reveal the critical behaviour of localization transitions and the critical exponents for the pure non-Hermitian AA and Stark limits, and for the non-Hermitian AAS model. We finally give a brief discussion and conclusion in Sec. \ref{secf}.

\section{\label{sec2} model and method}

We start by considering the combination of nonreciprocal hoppings and a linear gradient potential into the 1D quasiperiodic lattice of $L$ sites, as illustrated in Fig.~\ref{fig1} (a). The system can be described by the non-Hermitian AAS Hamiltonian:
\begin{equation}
\label{H}
\begin{aligned}
\hat{H}_{\text{nH-AAS}} = & -J\sum_{j=1}^{L-1}{(e^{-g}\hat{c}_{j}^\dagger \hat{c}_{j+1}+e^{g}\hat{c}_{j+1}^\dagger\hat{c}_{j}})  + \varepsilon \sum_{j=1}^{L} j \hat{c}_j^\dagger \hat{c}_j\\
& +W \sum_{j=1}^{L}\cos{(2\pi\alpha j+\phi)\hat{c}_j^\dagger \hat{c}_j}.
\end{aligned}
\end{equation}
Here $\hat{c}_j^\dagger$ ($\hat{c}_j)$ is the particle creation (annihilation) operator acting on site $j$, and $Je^{-g}=J_L$ ($Je^{g}=J_R$) represents the left- (right-) hopping strength between neighbor sites $j$ and $j+1$, with $g$ characterizing the non-Hermitian (nonreciprocal) strength. $\varepsilon$ is the gradient of Stark linear potential field. $W$ denotes the strength of the quasiperiodic potential, with $\alpha$ an irrational number and $\phi$ a random lattice phase uniformly chosen from the interval $[0,2\pi)$. We choose $\alpha$ as the inverse golden mean $\alpha = (\sqrt{5}-1)/2=\lim_{n\rightarrow\infty}F_n/F_{n+1}$ to approach an incommensurate lattice via two consecutive Fibonacci sequences $F_n$ and $F_{n+1}=L$. For convenience, we focus on the system under the OBC in exact diagonalization numerical calculations, and verify that our results preserve under the PBC. Hereafter, we set $J=1$ as the energy unit.

For $W=0$, the model Hamiltonian in Eq. (\ref{H}) returns to the pure non-Hermitian Stark model \cite{PhysRevLett.131.010801}, with Stark localization transition occurring when $\varepsilon>\varepsilon_{c}(g)$. For a vanishing Stark potential with $\varepsilon=0$, this model reduces to the non-Hermitian AA model with the critical localization point at $W_{c}(g)$. All eigenstates are extended and localized for $W<W_{c}(g)$ and $W>W_{c}(g)$ without mobility edges, respectively. In the thermodynamic limit, it has been revealed that the critical points $\varepsilon_c(g)=\varepsilon_c=0$ \cite{van_nieuwenburg_bloch_2019,PhysRevLett.131.010801} and $W_{c}(g)=2Je^{g}$ \cite{Jiang2019}, which are shown in Fig.~\ref{fig1} (b). We define $\delta=W-W_c(g)$ and illustrate the quantum criticality of the non-Hermitian AAS model in the $\delta$-$\varepsilon$ plane in Fig.~\ref{fig1} (b). Near the localization critical point at $\delta=\varepsilon=0$, one has the critical region A for the non-Hermitian AAS model. For $-W_c(g)< \delta < 0$ and infinitesimal small $\varepsilon$, there emerges the critical region B for the pure Stark localization. Since there exists no mobility edge in our model, we focus on the localization transition of the ground state and explore the quantum criticality. Hereafter, we denote critical exponents of merely quasiperiodicity-induced localization by subscript $\delta$, merely Stark-potential-induced localization by subscript $\varepsilon$, and both two effects with no subscript.

For finite-size systems, the wave function of the ground state is neither fully localized nor extended in the critical region, which rises up the critical phenomenon. As the system size increases, the critical region shrinks and finally collapses to the critical point in the thermodynamic limit.  Quantum phase transitions and corresponding critical behaviors can be characterized by several physical quantities respect to Hamiltonian parameters or system size. We adopt three characteristic quantities to investigate the localization-delocalization transition criticality in the non-Hermitian AAS model.

The first physical quantity is the localization length. In the critical region for both Hermitian and non-Hermitian cases, the localization length can be simply defined by a diverging length scale \cite{Sinha2019,S.Yin2022a,S.Yin2022b,S.Yin2022c,ljzhai2022,van_nieuwenburg_bloch_2019,PhysRevLett.131.010801}
\begin{equation}
\label{Eq:xiscaling}
\xi = \sqrt{\sum_{j=1}^{L} [( j - j_c )^2 ] |\psi(j)|^2},
\end{equation}
which actually denotes the width of a wave packet. Here $\psi(j)$ denotes the normalized wave function of the ground right eigenstate at site $j$, and $j_c\equiv\sum j|\psi(j)|^2$ is the localization center. In the critical region, $\xi$ scales as a power-law function of the distance between the parameter $\eta$ and the critical point $\eta_c(g)$:
\begin{equation}
	\label{Eq:xiscaling1}
	\xi\propto |\eta-\eta_c(g)|^{-\nu} =
	\begin{cases}
		\begin{split}
			& |W-W_{c} (g)|^{-\nu} & \quad  \varepsilon=0; \\
			& |\varepsilon-\varepsilon_{c}(g)|^{-\nu} & \quad \varepsilon \neq 0.
		\end{split}
	\end{cases}
\end{equation}
Here $\nu$ is the critical exponent, and the Hamiltonian parameter is $\eta = W$ for the pure AA model and $\eta = \varepsilon$ for the presence of the Stark potential. The second quantity used in our study is the inverse participation ratio (IPR) $\mathcal{I}$, which is defined as
\begin{equation}
\label{Eq:ipr}
{\mathcal{I} } = \frac{{\sum_{j=1}^L|\psi(j)|^4}}{\left({\sum_{j=1}^L|\psi(j)|^2}\right)^2}.
\end{equation}
Note that ${\mathcal{I}}$ scales as ${\mathcal{I}}\propto L^0$ for a localized state, while ${\mathcal{I}}\propto L^{-1}$ for an extended state. The critical behavior of ${\mathcal{I}}$ at the critical point satisfies the following scaling relation with respect to the system size
\begin{equation}
\label{Eq:iprscaling1}
{\mathcal{I} }\propto L^{-s/\nu},
\end{equation}	
with another critical exponent $s$ and the one previously mentioned $\nu$. When $L\rightarrow\infty$, ${\mathcal{I}}$ scales with the parameter distance $|\eta-\eta_c|$ as
\begin{equation}
	\label{Eq:iprscaling2}
	\mathcal{I}\propto |\eta-\eta_c(g)|^{s} =
	\begin{cases}
		\begin{split}
			&|W-W_{c} (g)|^{s} & \quad  \varepsilon=0; \\
			&|\varepsilon-\varepsilon_{c}(g)|^{s} & \quad \varepsilon \neq 0.
		\end{split}
	\end{cases}
\end{equation}
The final quantity used to characterize the quantum criticality is the energy gap between the ground state and first excited state. According to the finite-size scaling, the energy gap $\Delta E$ at the localization critical point scales as a power-law form
\begin{equation}
\label{Eq:gapscaling1}
   \Delta E \propto L^{-z},
\end{equation}
with $z$ being the third critical exponent. When $L\rightarrow\infty$, $\Delta E$ scales with the parameter distance as
\begin{equation}
	\label{Eq:gapscaling2}
	\Delta E\propto |\eta-\eta_c(g)|^{\nu z} =
	\begin{cases}
		\begin{split}
			&|W-W_{c} (g)|^{\nu z} & \quad  \varepsilon=0; \\
			&|\varepsilon-\varepsilon_{c}(g)|^{\nu z} & \quad \varepsilon \neq 0.
		\end{split}
	\end{cases}
\end{equation}

To explore the critical properties of the localization transition in the non-Hermitian AAS model, we use the finite-size scaling analysis and numerically extract the corresponding critical exponents in different situations, which are summarized in Fig.~\ref{fig1} (c). The scaling analysis takes the following ansatz
\begin{equation}
\label{Eq:generalscaling1}
   P(|\eta-\eta_c(g)|) = L^{\rho/\nu} f\left(|\eta-\eta_c(g)| L^{1/\nu}\right),
\end{equation}
where $P$ indicates the physical quantity $P\in\{\xi,\mathcal{I},\Delta E\}$, $f(.)$ is a universal scaling function, and $\rho \in \{\nu, -s, -\nu z \}$ is the critical exponent corresponding to the chosen physical quantity. In the thermodynamic limit $L\rightarrow\infty$, this finite-size ansatz recovers the scaling relation
\begin{equation}
\label{Eq:generalscaling}
\ P(\eta) \propto |\eta-\eta_c(g)|^{-\rho},
\end{equation}
as we introduced previously. Based on the scaling theory, we have numerically verified that the system exhibits the same critical exponents when approaching the critical point $\eta_c$ ($\eta=W,\varepsilon$) from both sides. When performing the finite-size scaling to extract critical exponents, the numerical data are taken from the right side of the critical point, i.e., $\eta>\eta_c$. In addition, we use the theoretical values of critical points $W_{c}(g)=2Je^{g}$ \cite{Jiang2019} and $\varepsilon_c(g)=0$ \cite{van_nieuwenburg_bloch_2019,PhysRevLett.131.010801} in our numerical simulations. We numerically verify that the finite-size critical points [$W_{c}^{(L)}$ and $\varepsilon_c^{(L)}$] are very close to the theoretical ones since the system size $L$ is sufficiently large.

\section{\label{sec3}Quantum criticality}
In this section, we aim to unveil the non-Hermitian effect on the localization transition criticality in the model. To this end, we first analyze the pure AA and Stark models with different non-Hermiticity strengths, with the corresponding critical exponents $\{\nu_{\delta},s_{\delta},z_{\delta}\}$ and $\{\nu_{\varepsilon},s_{\varepsilon},z_{\varepsilon}\}$, respectively. Then we combine the quasiperiodic with Stark potentials, and perform the scaling analysis to reveal new critical exponents, as summarized in Fig.~\ref{fig1} (c). At the end of this section, we further investigate the hybrid scaling in the overlap of critical regions A and B.

\subsection{Pure non-Hermitian AA and Stark criticalities}

\begin{figure}[t!]
	\centering
	\includegraphics[width=0.48\textwidth]{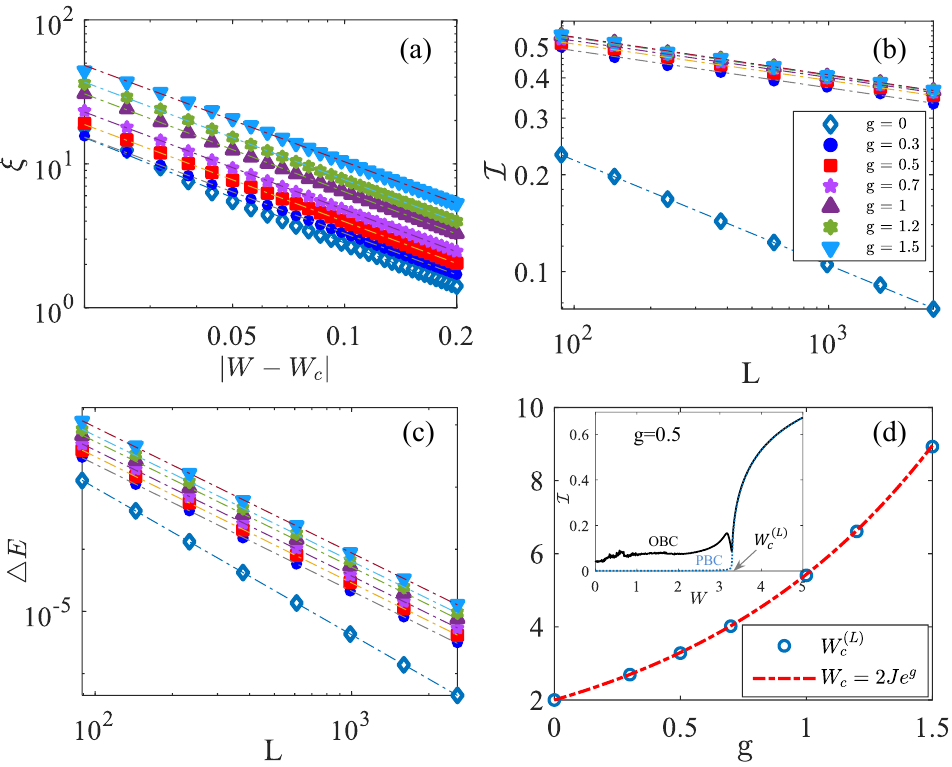}
	\caption{Scaling analysis in the pure non-Hermitian AA model with $\varepsilon=0$. Log-log plot of (a) localization length $\xi$ versus $|W-W_{c}|$, (b) IPR $\mathcal{I} $ and (c) energy gap $\Delta E $ at the critical point $W_{c}(g)=2Je^{g}$ \cite{Jiang2019} versus system size $L$. Dashed lines are linear fitting yielding exponents $\nu_{\delta}\approx \{1.0, 0.96\}$, $s_{\delta}\approx \{0.333, 0.12\}$, and $z_{\delta}\approx \{2.374, 2.0\}$ for $g=0$ (Hermitian limit) and $g\neq0$ (non-Hermitian cases), respectively. These obtained critical exponents are consistent with those reported in Refs. \cite{ljzhai2022,ewliang2024,Sinha2019}. (d) The extracted finite-size critical point $W_{c}^{(L)}$ (blue circles) and the theoretical one $W_{c}$ (red dashed line) versus $g$. The insert shows $W_{c}^{(L)}\approx3.3$ for $g=0.5$, which is extracted from the steep rise (the deep dive) of the IPR under the PBC (the OBC). All results are averaged over 1000 random $\phi$'s, and $L=987$ is used in (a-d).}
	\label{fig2}
\end{figure}

The model Hamiltonian $\hat{H}_{\text{nH-AAS}}$ with $\varepsilon=0$ recovers to the pure non-Hermitian AA model with the localization transition at $W_{c}(g)=2Je^{g}$ in the thermodynamic limit \cite{Jiang2019}. Near this critical point, we use the scaling relations in Eqs. (\ref{Eq:xiscaling1},\ref{Eq:iprscaling1},\ref{Eq:gapscaling1}) with $\varepsilon=0$ to obtain the critical exponents $\{\nu_{\delta},s_{\delta},z_{\delta}\}$ labeled by the subscript $\delta$, as shown in Fig.~\ref{fig2}. The localization length $\xi$ versus $|W-W_{c}(g)|$ for different nonreciprocal strengths are shown in Fig.~\ref{fig2} (a) with a double-log axes. The scaling relation in Eq. (\ref{Eq:xiscaling1}) is clearly revealed, and the linear fit of the critical exponent $\nu_{\delta}\approx1$ for $g=0$ and $\nu_{\delta}\approx0.96$ for $g\neq 0$, which are consistent with those approximately obtained in Ref. \cite{ljzhai2022}. Moreover, the results for different values of $g$ are displayed in Fig.~\ref{fig2} (a) suggest that $\nu_{\delta}$ is a universal exponent independent of $g$ for non-Hermitian models, and is different from that in the Hermitian case. Note that the localization lengths approaching the critical point $W_c$ from both sides share the same critical exponent \cite{ljzhai2022} under both the OBC and the PBC, although they are different in most regions of two sides $W>W_c$ [Fig.~\ref{fig2} (a)] and $W<W_c$ (not shown here). Fig.~\ref{fig2} (b) shows the IPR $\mathcal{I}$ versus $L$ at the localization transition point satisfies Eq. (\ref{Eq:iprscaling1}), with the critical exponent $s_{\delta}\approx 0.33$ for $g = 0$ and $s_{\delta}\approx 0.12$ for $g \neq 0$, respectively. We plot the energy gap $\Delta E$ at the critical point versus $L$ in Fig.~\ref{fig2} (c), which agrees with Eq. (\ref{Eq:gapscaling1}) with $z_{\delta}\approx2.374$ for $g=0$ and $z_{\delta}\approx2$ for $g \neq 0$, respectively. The finite-size critical point $W_{c}^{(L)}$ with respect to $g$ for $L=987$ is shown in Fig.~\ref{fig2} (d), which is fitted well by the theoretical value $W_{c}=2Je^{g}$. Here $W_{c}^{(L)}$ is identically extracted from the steep rise of the IPR $\mathcal{I}$ under the PBC and the deep dive of $\mathcal{I}$ under the OBC~\cite{Jiang2019}, with an example shown in the insert of Fig.~\ref{fig2} (d). The extended state for $W<W_{c}^{(L)}$ under the PBC becomes a localized skin mode under the OBC, while the AL state for $W>W_{c}^{(L)}$ is insensitive to the boundaries. Thus, the onset of AL at $W=W_{c}^{(L)}$ under both boundary conditions are identical. All numerically obtained critical exponents $\{\nu,s,z\}=\{\nu_{\delta},s_{\delta},z_{\delta}\}$ for the pure AA model are summarized in Fig. \ref{fig1} (c) with Hermitian in the first column and its non-Hermitian counterpart in the second column, which are consistent with those reported in Refs. \cite{ljzhai2022,ewliang2024,Sinha2019}.

\begin{figure}[t!]
	\centering
	\includegraphics[width=0.48\textwidth]{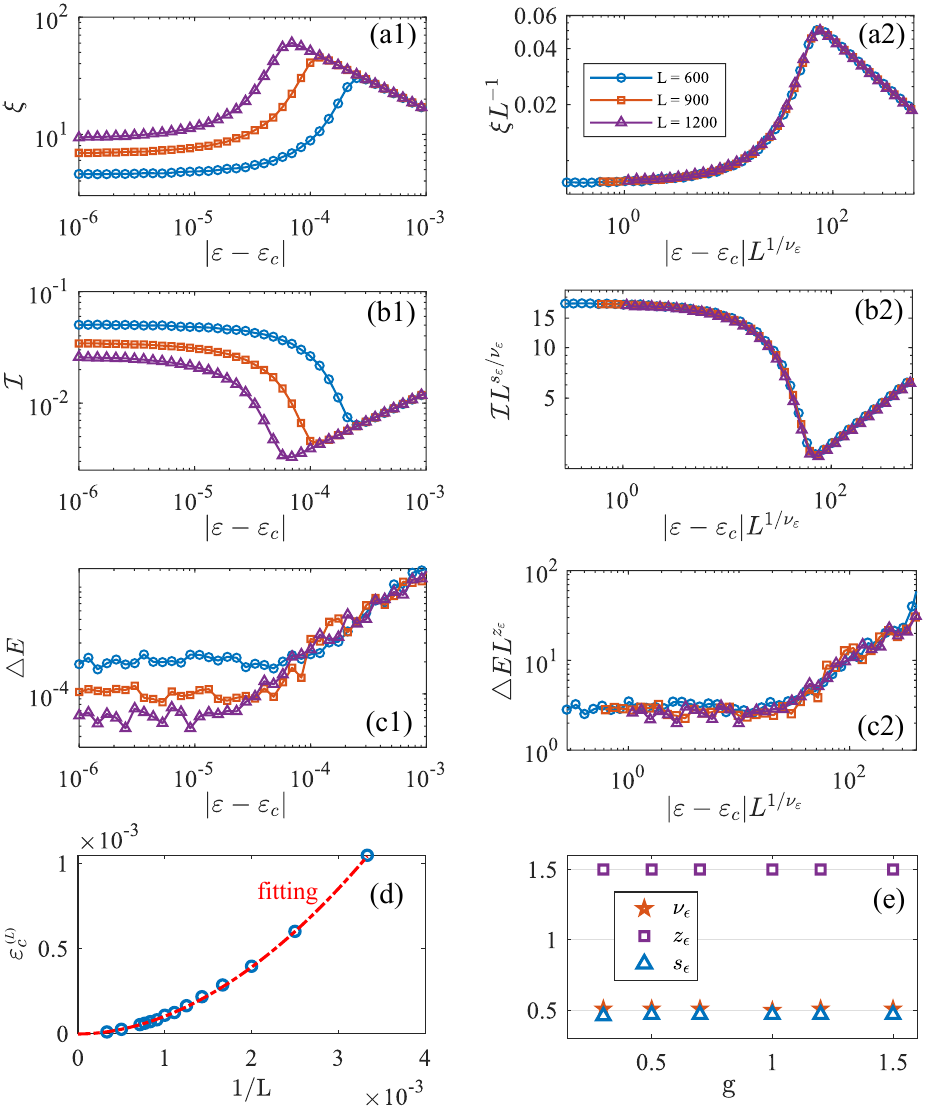}
	\caption{Scaling analysis in the pure non-Hermitian Stark model with $W=0$. (a1, a2) Log-log plot of $\xi$ versus $|\varepsilon-\varepsilon_{c}|$ before (a1) and after (a2) rescaling for different $L$. (b1, b2) Log-log plot of $\mathcal{I}$ versus $|\varepsilon-\varepsilon_{c}|$ before (b1) and after (b2) rescaling. (c1, c2) Log-log plot of $\Delta E $ versus $|\varepsilon-\varepsilon_{c}|$ before (c1) and after (c2) rescaling. (d) The finite-size critical point $\varepsilon_{c}^{(L)}$ versus $1/L$. Here $\varepsilon_{c}^{(L)}$ is identically extracted from the maximum value of $\xi$ and the minimum value of $\mathcal{I}$, and the red dashed line denotes the best fitting. (e) Critical exponents versus $g$ for the pure non-Hermitian Stark model show that $\nu_{\varepsilon}\approx 0.51$, $s_{\varepsilon}\approx 0.47$, and $z_{\varepsilon}\approx 1.5$ for all $g\neq0$. $g=0.5$ is used in (a-d).}
	\label{fig3}
\end{figure}

In the pure non-Hermitian Stark model when $W=0$, the localization transition occurs at $\varepsilon=\varepsilon_{c}$ with $\varepsilon_{c}=0$ in the thermodynamic limit [see Fig.~\ref{fig3}(d)]. In this case, we use the finite-size scaling of physical quantities $\{\xi,\mathcal{I},\Delta E\}$ to reveal the critical exponents $\{\nu_{\varepsilon},s_{\varepsilon},z_{\varepsilon}\}$ labeled by subscript $\varepsilon$. As critical behaviors are independent of $g$ as long as $g\neq0$, we first discuss the results for $g=0.5$ and then numerically confirm the independence of the critical exponents on $g$ [see Fig.~\ref{fig3}(e)]. Notably, the scaling functions of $\{\xi,\mathcal{I},\Delta E\}$ for the Hermitian Stark model with $g=0$ has been obtained with $\{\nu_{\varepsilon},s_{\varepsilon},z_{\varepsilon}\}\approx\{0.33,0.33,2\}$ in Ref. \cite{ewliang2024}, similar as those in Ref. \cite{PhysRevLett.131.010801}. Here we generalize the scaling analysis to the non-Hermitian Stark model with $g\neq0$ and obtain new critical exponents $\{\nu_{\varepsilon},s_{\varepsilon},z_{\varepsilon}\}\approx\{0.51,0.47,1.5\}$, as summarized in Fig.~\ref{fig1}(c).

The finite-size scaling form of the localization length $\xi$ can be derived from Eq. (\ref{Eq:generalscaling1}) for $W=0$ as
\begin{equation}
\label{Eq:xiscaling3}
\xi=L f_1\left(|\varepsilon-\varepsilon_{c}(g)| L^{1/\nu_\varepsilon}\right).
\end{equation}
Here $f_1(\cdot)$ [and $f_i(\cdot)$] denotes a universal function where all data points collapse onto after rescaling. Our numerical results of $\xi$ as functions of $|\varepsilon-\varepsilon_{c}(g)|$ for various $L$'s are shown in Fig.~\ref{fig3} (a1). The enhancement of $\xi$ as increasing $\varepsilon$ is due to the localization of the wave function caused by the non-Hermitian skin effect in finite-size systems under the OBC. It is suppressed by the Stark localization as $|\varepsilon-\varepsilon_{c}(g)|$ is increased. Over a certain threshold of $|\varepsilon-\varepsilon_{c}(g)|$, the localization length becomes independent on the system size $L$. This indicates the emergence of Stark localization on the finite systems with $\varepsilon_{c}$ \cite{PhysRevLett.131.010801}. By rescaling $x$ axis $|\varepsilon-\varepsilon_{c}(g)|$ and $y$ axis $\xi$ as $|\varepsilon-\varepsilon_{c}(g)| L^{1/\nu_{\varepsilon}}$ and $\xi L^{-1}$ in Fig.~\ref{fig3} (a2), we find the best collapse for our numerical data by choosing $\nu_{\varepsilon}=0.51$. Similarly, the IPR $\mathcal{I}$ in this situation satisfies the scaling relation
\begin{equation}
\label{Eq:iprscaling4}
{\mathcal{I} }=L^{-s_\varepsilon/\nu_\varepsilon} f_2\left(|\varepsilon-\varepsilon_{c}(g)| L^{1/\nu_\varepsilon}\right).
\end{equation}
Figure \ref{fig3} (b1) shows the numerical results of $\mathcal{I}$ versus $|\varepsilon-\varepsilon_{c}(g)|$ for several $L$'s. Similar as the localization length, a small value of $\mathcal{I}$ away from zero exhibits for all curves due to the skin-effect induced localization. Beyond a certain threshold when the ground state is Stark localized, the curves become independent of $L$. We rescale $\mathcal{I}$ and $|\varepsilon-\varepsilon_{c}(g)|$ as $\mathcal{I} L^{s_{\varepsilon}/v_{\varepsilon}}$ and $|\varepsilon-\varepsilon_{c}(g)| L^{1/\nu_{\varepsilon}}$, and obtain the critical exponent $s_{\varepsilon} = 0.47$ from the IPR, as shown in Fig.~\ref{fig3} (b2). For the energy gap $\Delta E$, we adopt the following scaling form
\begin{equation}
\label{Eq:gapscaling4}
\Delta E=L^{-z_\varepsilon} f_3\left(|\varepsilon-\varepsilon_{c}(g)| L^{1/\nu_\varepsilon}\right).
\end{equation}
In Fig. \ref{fig3} (c1), we depict $\Delta E$ versus the distance $|\varepsilon-\varepsilon_{c}(g)|$ for different lattice sizes. With the increasing of the distance, the energy gap first shows the size-dependence and then becomes size-independent after the Stark localization transition, which is consistent with the results of $\xi$ and $\mathcal{I}$. The critical exponent $z_\varepsilon$ is then determined through the data collapses in Fig.~\ref{fig3} (c2), which yields the best fitting $z_\varepsilon=1.5$. Note that the critical point for the Stark localization in finite-size systems $\varepsilon_{c}^{(L)}$ can be identically extracted from the maximum value of $\xi$ and the minimum value of $\mathcal{I}$ \cite{PhysRevLett.131.010801}, i.e., $\varepsilon_{c}^{(L)}=\max[\xi(\varepsilon)]=\min[\mathcal{I}(\varepsilon)]$ [see Figs.~\ref{fig3} (a1) and (b1)]. We show the extracted finite-size critical point $\varepsilon_{c}^{(L)}$ versus $1/L$ in Fig.~\ref{fig3} (d). One can see that $\varepsilon_{c}^{(L)}\rightarrow\varepsilon_{c}=0$ in the thermodynamic limit $L\rightarrow\infty$, and the values of $\varepsilon_{c}^{(L)}$ are already very close to zero when $L=600,900,1200$ are taken in Fig. \ref{fig3}. Finally, we show the extracted critical exponents $\{\nu_{\varepsilon},s_{\varepsilon},z_{\varepsilon}\}$ versus the non-Hermitian parameter $g$ in Fig.~\ref{fig3}(e) and find that they preserve for all $g\neq0$.

\subsection{\label{B} Non-Hermitian AAS criticality}

We have revealed the non-Hermitian effects on the critical behaviors of the pure AA and Stark models. Now we proceed to investigate the critical properties in the non-Hermitian AAS model by analyzing the effect of the Stark potential on the non-Hermitian AA critical point $W=W_{c}(g)$. We first numerically calculate the physical quantities $\{\xi,\mathcal{I},\Delta E\}$ for varying $\eta=|\varepsilon-\varepsilon_{c}(g)|$ and fixed $W=W_{c}(g)$ with $g=0.5$. This enables us to obtain the simplified scaling functions [see Eqs. (\ref{Eq:xiscaling4},\ref{Eq:iprscaling5},\ref{Eq:gapscaling5})] with corresponding critical exponents $\{\nu,s,z\}\approx\{0.33,0.038,2\}$. They are verified to be the same for the non-Hermitian AAS model with all $g\neq0$, but different from the counterparts $\{\nu,s,z\}\approx\{0.3,0.098,2.374\}$ for the Hermitian AAS model obtained in Ref. \cite{ewliang2024}, as summarized in Fig.~\ref{fig1} (c). We then extend the scaling analysis to the whole critical region and obtain the generalized scaling functions [see Eqs.~(\ref{Eq:xiscaling5},\ref{Eq:iprscaling6},\ref{Eq:gapscaling6})] with two scaling variables and the same critical exponents. Moreover, the constraint on the scaling functions in the overlap critical region leads to a hybrid scaling form with a hybrid exponent [see Eq.~(\ref{Eq:xiscaling7})].

\begin{figure}[t!]
	\centering
	\includegraphics[width=0.48\textwidth]{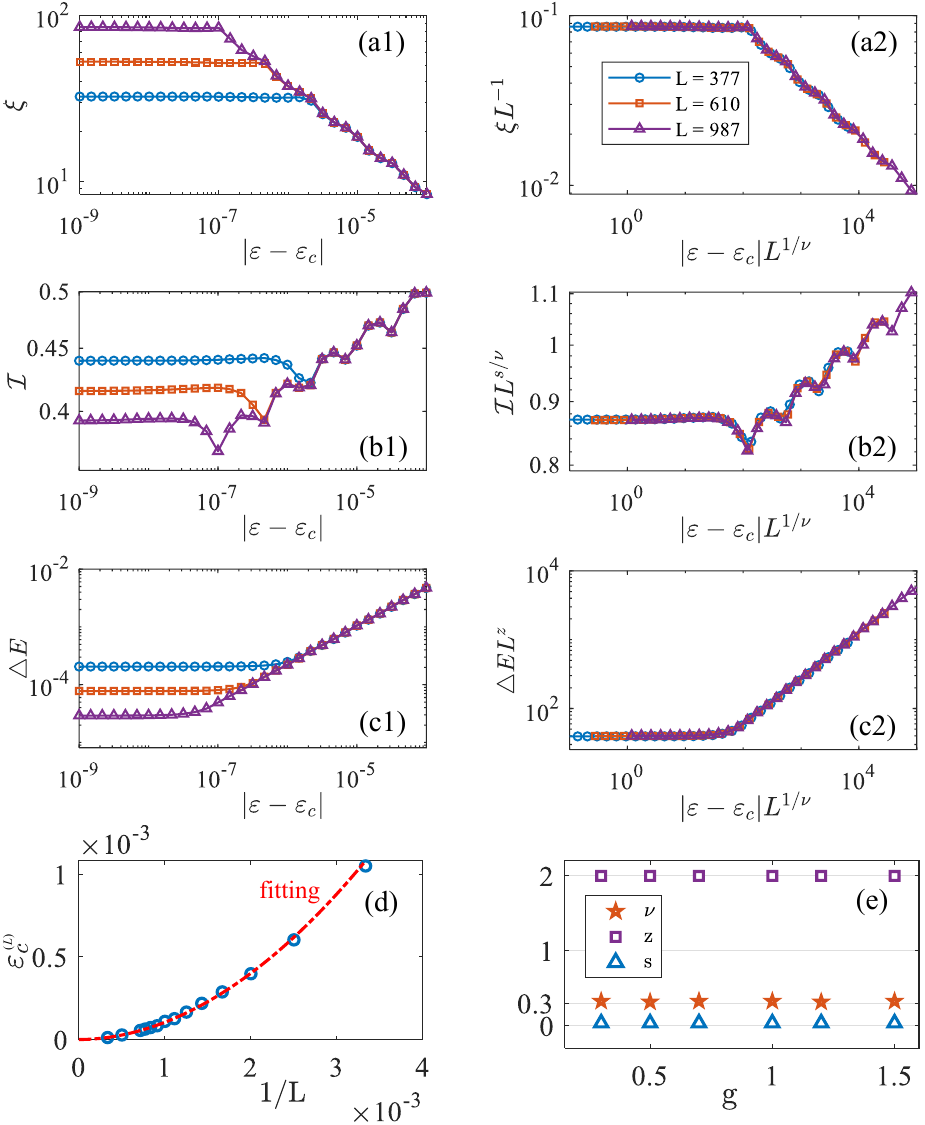}
	\caption{Scaling analysis in the non-Hermitian AAS model at $W=W_{c}(g)$. (a1, a2) Log-log plot of $\xi$ versus $|\varepsilon-\varepsilon_{c}|$ before (a1) and after (a2) rescaling for different $L$. (b1, b2) Log-log plot of $\mathcal{I} $ versus $|\varepsilon-\varepsilon_{c}|$ before (b1) and after (b2) rescaling. (c1, c2) Log-log plot of $\Delta E $ versus $|\varepsilon-\varepsilon_{c}|$ before (c1) and after (c2) rescaling. (d) The numerically extracted critical point for finite-size systems $\varepsilon_{c}^{(L)}$ versus $1/L$ (blue circles) with the best fitting (red dashed line). (e) Critical exponents versus $g$ for the non-Hermitian AAS model show that $\nu\approx 0.33$, $s\approx 0.038$ and $z\approx 2$ for all $g\neq0$. The results are averaged over 1000 random $\phi$'s, and $g=0.5$ is used in (a-d).}
	\label{fig4}
\end{figure}

\begin{figure}[t!]
	\centering
	\includegraphics[width=0.48\textwidth]{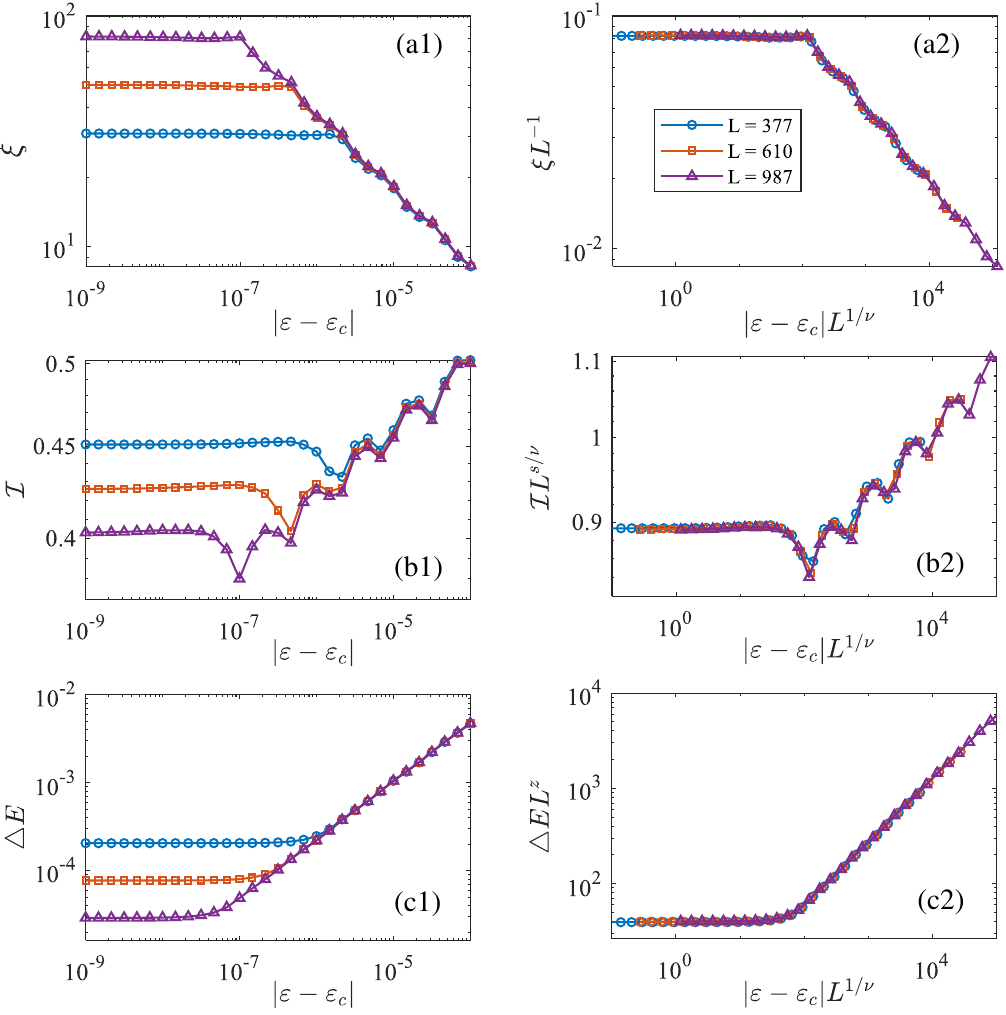}
	\caption{Scaling analysis in the non-Hermitian AAS model with fixed $(W-W_{c}(g)) L^{1/\nu_{\delta}}=1$. (a1, a2) Log-log plot of $\xi$ versus $|\varepsilon-\varepsilon_{c}|$ before (a1) and after (a2) rescaling for different system sizes $L$. (b1, b2) Log-log plot of $\mathcal{I} $ versus $|\varepsilon-\varepsilon_{c}|$ before (b1) and after (b2) rescaling. (c1, c2) Log-log plot of $\Delta E $ versus $|\varepsilon-\varepsilon_{c}|$ before (c1) and after (c2) rescaling. Results are averaged over 1000 $\phi$'s, and $g$=0.5 is used in (a-c).}
	\label{fig5}
\end{figure}

\begin{figure}[t!]
	\centering
	\includegraphics[width=0.48\textwidth]{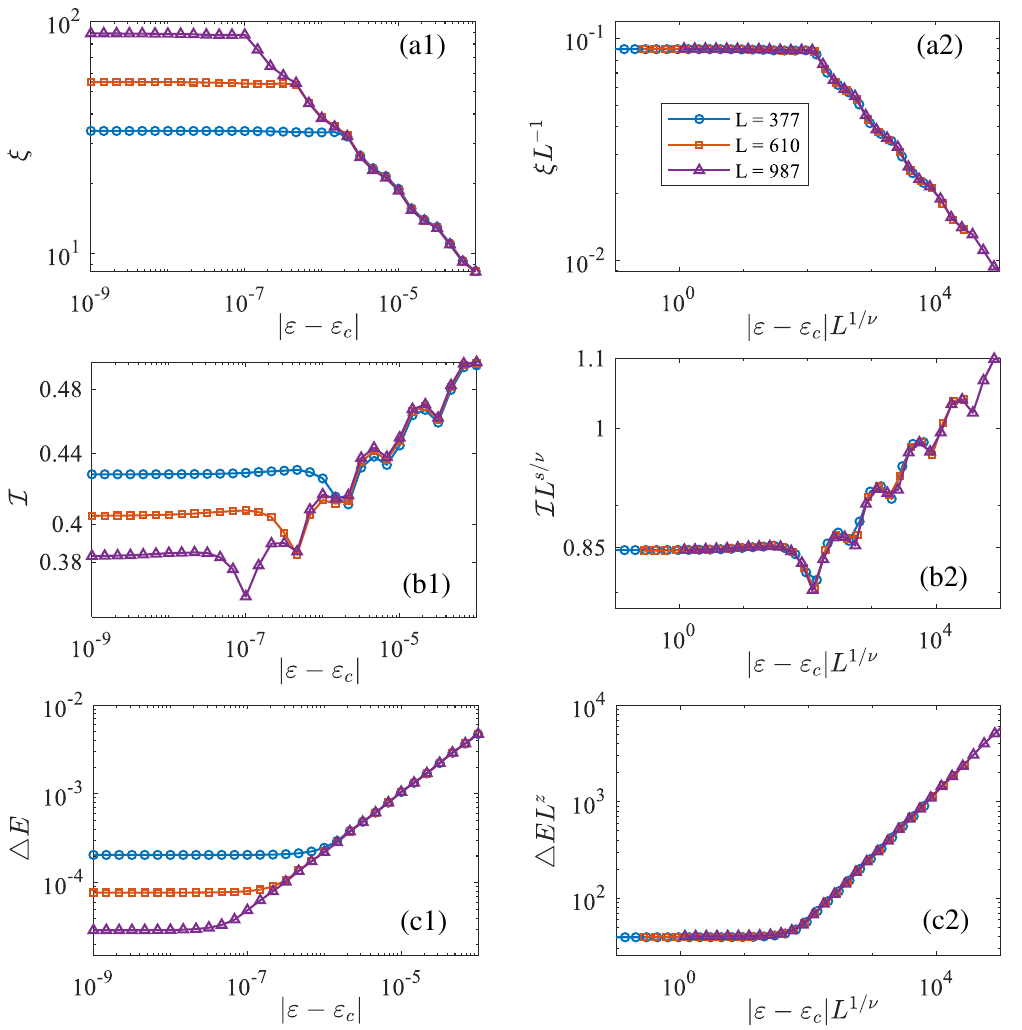}
	\caption{Scaling analysis in the non-Hermitian AAS model with fixed $(W-W_{c}(g)) L^{1/\nu_{\delta}}=-1$. (a1, a2) Log-log plot of $\xi$ versus $|\varepsilon-\varepsilon_{c}|$ before (a1) and after (a2) rescaling for different system sizes $L$. (b1, b2) Log-log plot of $\mathcal{I} $ versus $|\varepsilon-\varepsilon_{c}|$ before (b1) and after (b2) rescaling. (c1, c2) Log-log plot of $\Delta E $ versus $|\varepsilon-\varepsilon_{c}|$ before (c1) and after (c2) rescaling. Results are averaged over 1000 $\phi$'s, and $g$=0.5 is used in (a-c).}
    \label{fig6}
\end{figure}

By setting $W=W_{c}(g)$, one has $\delta=W-W_{c}=0$ and varies $\varepsilon$ in the critical region along the y axis in Fig.~\ref{fig1}(b). In this case, we obtain the finite-size scaling functions of the three physical quantities with respect to the distance $|\varepsilon-\varepsilon_{c}(g)|$ and the system size $L$ as
\begin{align}
\label{Eq:xiscaling4} &\xi=Lf_4\left(|\varepsilon-\varepsilon_{c}(g)| L^{1/\nu}\right),\\
\label{Eq:iprscaling5}&{\mathcal{I} }=L^{-s/\nu} f_5\left(|\varepsilon-\varepsilon_{c}(g)| L^{1/\nu}\right), \\
\label{Eq:gapscaling5}&\Delta E=L^{-z} f_6\left(|\varepsilon-\varepsilon_{c}(g)| L^{1/\nu}\right).
\end{align}
The critical exponents for the non-Hermitian AAS model can be numerically determined from the collapse of rescaled data. Firstly, we numerically obtain localization length $\xi$ versus the distance $|\varepsilon-\varepsilon_{c}(g)|$ for several system sizes, as shown in Fig.~\ref{fig4} (a1). All curves become size-independent after crossing the critical point. According to Eq.~(\ref{Eq:xiscaling4}), we rescale $\xi$ and $|\varepsilon-\varepsilon_{c}(g)|$ as $\xi L^{-1}$ and $|\varepsilon-\varepsilon_{c}(g)| L^{1/\nu}$ and plot them in Fig.~\ref{fig4} (a2), where all curves collapse into a single one with $\nu=0.33$. Apparently, the critical exponent $\nu=0.33$ is different from $\nu_{\delta}=0.96$ and $\nu_{\varepsilon}=0.51$ in the pure non-Hermitian AA and Stark limits, as well as $\nu=0.3$ for the Hermitian AAS model \cite{ewliang2024}. This result implies the Stark potential can contributes a new relevant direction at the non-Hermitian AA critical point at $W=W_{c}(g)$. It's worth noticing that $\nu\approx\nu_{\delta}/3$ indicates that the Stark potential exhibits a short-range correlation. Since the quasiperiodic potential is known as infinitely correlated, the Stark potential is less relevant in the critical localization region. Figure \ref{fig4} (b1) shows the IPR $\mathcal{I}$ versus $|\varepsilon-\varepsilon_{c}(g)|$ for different system sizes, with the size-dependent transition point $\varepsilon_{c}$. In Fig.~\ref{fig4}(b2), we rescale $\mathcal{I}$ and $|\varepsilon-\varepsilon_{c}(g)|$ as $\mathcal{I} L^{s/\nu}$ and $|\varepsilon-\varepsilon_{c}(g)| L^{1/\nu}$ according to Eq.~(\ref{Eq:iprscaling5}), where the best collapse of all curves is achieved by setting exponent $s=0.038$. Again, the critical exponent is distinct from $s_{\delta}=0.11$ and $s_{\varepsilon}=0.47$ for the pure non-Hermitian AA and Stark limits. Notably, the ratio $s/\nu\approx s_{\delta}/\nu_{\delta}\approx0.115$, which indicates that at the non-Hermitian AA criticality $W=W_c(g)$, the finite-size scaling of the IPR given by Eq. (\ref{Eq:iprscaling1}) preserves in the presence of the Stark potential. We present the energy gap $\Delta E$ as functions of $|\varepsilon-\varepsilon_{c}(g)|$ for various system sizes in Fig. \ref{fig4} (c1), and the rescaled curves according to Eq.~(\ref{Eq:gapscaling5}) in Fig. \ref{fig4} (c2). The size dependence of the energy gap is consistent with those found in localization length and the IPR. The critical exponent $z=2$ obtained from the data collapse is the same as that in the pure non-Hermitian AA model $z_{\delta}=2$ for $g\neq0$. When $g=0$, the same critical exponent $z=2.374$ for Hermitian AA and AAS models has been obtained in Ref. \cite{ewliang2024}. Furthermore, the finite-size critical point $\varepsilon_{c}^{(L)}$ versus $1/L$ is plotted in Fig.~\ref{fig4} (d), which shows that $\varepsilon_{c}^{(L)}\rightarrow\varepsilon_{c}=0$ when $L \rightarrow \infty$. Note that $\varepsilon_c^{(L)}$ is close to zero when we choose $L=377,610,987$ in Figs.~\ref{fig4} (a-c). Finally, we verify the non-Hermitian parameter $g$ and numerically extracted the corresponding values of $\{\nu,s,z\}$ in Fig.~\ref{fig4} (e), which very the same critical exponents in the non-Hermitian AAS model with $g\neq0$.

\begin{figure}[t!]
	\centering
	\includegraphics[width=0.48\textwidth]{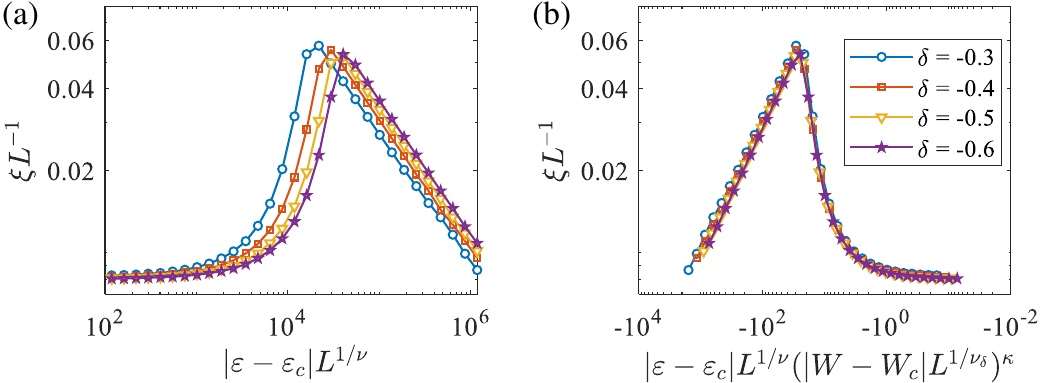}
	\caption{(a) Curves of $\xi L^{-1}$ versus $|\varepsilon-\varepsilon_{c}| L^{1/\nu}$ for various values of $\delta$ with fixed system size $L=987$ and $g=0.5$. (b) Curves of $\xi L^{-1}$ versus rescaled $|\varepsilon-\varepsilon_{c}| L^{1/\nu}(|W-W_{c}| L^{1/\nu_{\delta}})^{\kappa}$ where data for different $\delta$ collapse by setting the hybrid exponent $\kappa=-1.1$. The results are averaged over 1000 $\phi$'s.}
	\label{fig7}
\end{figure}

We further perform the scaling analysis for the general case of $W\neq W_{c}(g)$ in the critical region A [see Fig. \ref{fig1} (b)]. In this region, the critical localization behavior in the non-Hermitian AAS model depends on both two distances $|W-W_{c}(g)|$ and $|\varepsilon-\varepsilon_{c}(g)|$. Thus the scaling functions for $W=W_{c}(g)$ given by Eqs. (\ref{Eq:xiscaling4},\ref{Eq:iprscaling5},\ref{Eq:gapscaling5}) need to be modified. Concretely, the scaling behaviors of three adopted quantities $\{\xi, \mathcal{I}, \Delta E\}$ introduced in Sec.~\ref{sec2} are generalized to two scaling variables $|W-W_{c}(g)|$ and $|\varepsilon-\varepsilon_{c}(g)|$:
\begin{align}
	\label{Eq:xiscaling5} &\xi=L f_{7}\left(|W-W_{c}(g)| L^{1/\nu_\delta},|\varepsilon-\varepsilon_{c}(g)| L^{1/\nu}\right),\\
	\label{Eq:iprscaling6}&\mathcal{I} =L^{-s/\nu} f_{8}\left(|W-W_{c}(g)| L^{1/\nu_\delta},|\varepsilon-\varepsilon_{c}(g)| L^{1/\nu}\right), \\
	\label{Eq:gapscaling6}&\Delta E=L^{-z} f_{9}\left(|W-W_{c}(g)| L^{1/\nu_\delta},|\varepsilon-\varepsilon_{c}(g)| L^{1/\nu}\right),
\end{align}
Using these generalized scaling formulas, we can analytically derive the identity of the critical exponents $z=z_{\delta}$ and the exponent ratio $s/\nu=s_{\delta}/\nu_{\delta}$, which has been  numerically confirmed at the case of $W-W_{c}(g)=0$. The proof is similar to that in the Hermitian AAS model \cite{ewliang2024}. We first consider the IPR $\mathcal{I}$ in Eq. (\ref{Eq:iprscaling6}) for $|\varepsilon-\varepsilon_{c}(g)|=0$ and $L\rightarrow\infty$. This leads to the scaling $\mathcal{I} \propto |W-W_{c}(g)|^{s \nu_{\delta}/\nu}$. By comparing this equation with Eq.~(\ref{Eq:iprscaling2}), we can directly obtain $s/\nu =s_{\delta}/\nu_{\delta}$. We then consider $\Delta E$ in Eq. (\ref{Eq:gapscaling6}) for $|\varepsilon-\varepsilon_{c}(g)|=0$ and $L\rightarrow\infty$, which reduces to $\Delta E \propto |W-W_{c}(g)|^{\nu_{\delta}z}$. Along with scaling equation $\Delta E \propto |W-W_{c}(g)|^{\nu_{\delta}z_{\delta}}$ for the pure non-Hermitian AA criticality, one realizes $z=z_{\delta}$ for these two models. Finally, we  numerically compute the quantities $\{\xi, \mathcal{I}, \Delta E\}$ in the critical region for fixed $(W-W_{c}(g)) L^{1/\nu_{\delta}}$=1 ($\delta=W-W_{c}>0$) in Fig.~\ref{fig5} and $(W-W_{c}(g)) L^{1/\nu_{\delta}}=-1$ ($\delta<0$) in Fig.~\ref{fig6}, respectively. The results validate the faithfulness of the scaling forms given in Eqs.~(\ref{Eq:xiscaling5},\ref{Eq:iprscaling6},\ref{Eq:gapscaling6}). For both situations in Figs.~\ref{fig5} and \ref{fig6}, the collapse of rescaled curves according to Eqs.~(\ref{Eq:xiscaling5},\ref{Eq:iprscaling6},\ref{Eq:gapscaling6}) is achieved with the same critical exponents $\{\nu,s,z\}=\{0.33,0.038,2\}$ obtained in Fig. \ref{fig4} for the specific case of $W=W_{c}(g)$.

When the quasiperiodic potential strength $W-W_{c}<0$, there is an overlap between the critical regions A and B for the non-Hermitian AAS and Stark models, as shown in Fig. \ref{fig1} (b). A constraint on the scaling functions should be imposed in this overlap region, which gives rise to the hybrid scaling form~\cite{S.Yin2022a}. As a result, the critical behaviors is simultaneously described by the scaling forms of the non-Hermitian AAS and Stark models. For instance, the localization length $\xi$ should obey both scaling forms given in Eq.~(\ref{Eq:xiscaling3}) and Eq.~(\ref{Eq:xiscaling5}). This requirement suggests the hybrid scaling form of $\xi$:
\begin{equation}
    \label{Eq:xiscaling7}
    \xi = L f_{10} \left(|\varepsilon-\varepsilon_{c}(g)| L^{\frac{1}{\nu}} (|W-W_{c}(g)| L^{\frac{1}{\nu_{\delta}}})^{\kappa}\right),
 \end{equation}
where $\kappa=\nu_{\delta}(1/\nu_{\varepsilon}-1/\nu)=-1.1$ is a hybrid critical exponent. In Fig. \ref{fig7} (a), we depict the numerical results of $\xi L^{-1}$ versus rescaled variable $|\varepsilon-\varepsilon_{c}(g)| L^{1/\nu}$ for several $\delta$ and fixed system size $L=987$. By setting $\kappa = -1.1$ from the theoretical prediction, we find all curves collapse well onto one curve with respect to the hybrid quantity $|\varepsilon-\varepsilon_{c}(g)| L^{{1}/{\nu}} (|W-W_{c}(g)| L^{{1}/{\nu_{\delta}}})^{\kappa}$, as shown in Fig. \ref{fig7} (b). This confirms the hybrid scaling form of the localization length with the hybrid critical exponent given in Eq.~(\ref{Eq:xiscaling7}).

\begin{figure}[t!]
	\centering
	\includegraphics[width=0.48\textwidth]{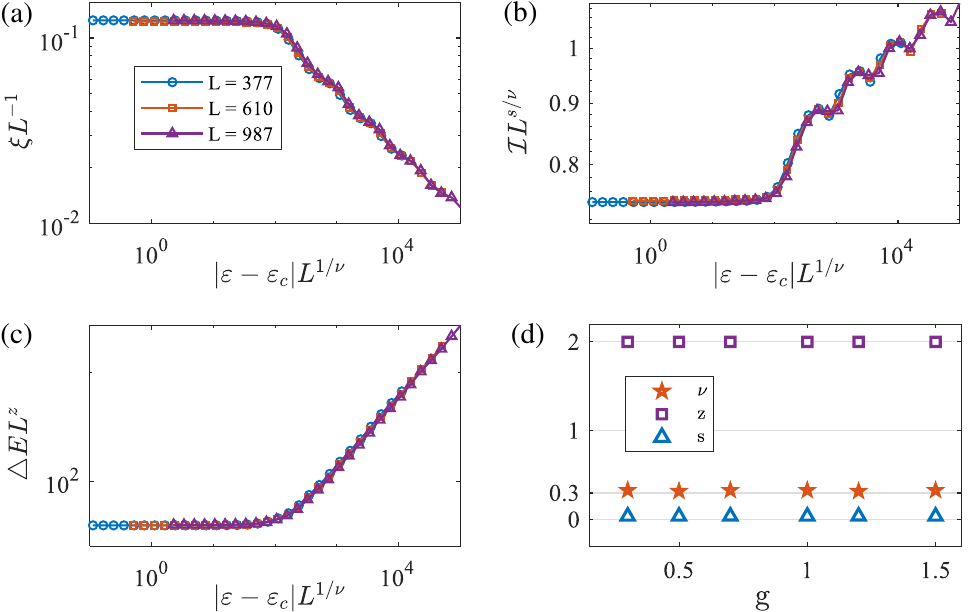}
	\caption{Scaling analysis in the non-Hermitian AAS model under the PBC at $W=W_{c}(g)$ for $g=0.5$. (a) Rescaled curves of $\xi L^{-1}$ versus $|\varepsilon-\varepsilon_{c}| L^{1/\nu}$ collapse onto each other for $\nu = 0.33$. (b) Rescaled $\mathcal{I} L^{s/\nu}$ versus $|\varepsilon-\varepsilon_{c}| L^{1/\nu}$ collapse for $s = 0.038$. (c) Rescaled $\Delta E L^{z} $ versus $|\varepsilon-\varepsilon_{c}| L^{1/\nu}$ collapse for $z = 2$. (d) Extracted critical exponents under the PBC versus $g$. All results are averaged over 1000 $\phi$'s.}
	\label{fig8}
\end{figure}

\section{\label{secf}Discussion and Conclusion}

Before concluding, we discuss the quantum criticality of localization transition in the non-Hermitian AAS model under the PBC. We consider the non-Hermitian strength $g=0.5$ and show that the scaling functions and critical exponents remain the same under both the PBC and the OBC. The finite-size scaling form for the physical quantities $\{\xi,\mathcal{I},\Delta E\}$ under the PBC are still given by Eqs.~(\ref{Eq:xiscaling4}, \ref{Eq:iprscaling5}, \ref{Eq:gapscaling5}). In this case, we rescale localization length $\xi$ and $|\varepsilon-\varepsilon_{c}(g)|$ as $\xi L^{-1}$ and $|\varepsilon-\varepsilon_{c}(g)| L^{1/\nu}$ respectively according to Eq.~(\ref{Eq:xiscaling4}), and show that all curves collapse into a single one with $\nu=0.33$ in Fig.~\ref{fig8} (a). Similarly, we rescale IPR $\mathcal{I}$ and $|\varepsilon-\varepsilon_{c}(g)|$ as $\mathcal{I} L^{s/\nu}$ and $|\varepsilon-\varepsilon_{c}(g)| L^{1/\nu}$ respectively according to Eq.~(\ref{Eq:iprscaling5}) in Fig.~\ref{fig8} (b), which confirms the collapse under the same critical exponent $s=0.038$ under the PBC. The rescale of the energy gap reads $\Delta E L^{z}$ and $|\varepsilon-\varepsilon_{c}(g)| L^{1/\nu}$, with the critical exponent $z=2$, as shown in Fig. \ref{fig8} (c). We plot the extracted critical exponents for various $g$ under the PBC in Fig. \ref{fig8} (d), which reveals the same results with those in Fig. \ref{fig4} (e) under the OBC. The same scaling functions and critical exponents, as well as their independence on the non-Hermitian parameter, indicate that the skin effect under the OBC does not affect the critical behaviour in the non-Hermitian AAS model.

In summary, we have explored the quantum criticality of the localization-delocalization transition in the non-Hermitian AAS model. We have derived and demonstrated several scaling functions for the localization length, the IPR and the energy gap in different critical regions. With the scaling functions, we have numerically obtained different critical exponents for the non-Hermitian AAS model and in the pure non-Hermitian AA and Stark limits, as presented in Fig. \ref{fig1} (c) with two groups of newly revealed critical exponents. These critical exponents in the non-Hermitian situations are totally distinct to their Hermitian counterparts. We have also found that although the scaling functions are relevant to the non-Hermitian parameter, the critical exponents are independent of the non-Hermitian strength and remain the same under different boundary conditions. Moreover, we have revealed the hybrid scaling function with a hybrid critical exponent in the overlap critical region for the non-Hermitian AAS and Stark models. The symmetry analysis and classification of the localization transitions in the non-Hermitian AAS and disordered AA models require a further investigation.

\begin{acknowledgments}
This work was supported by the National Natural Science Foundation of China (Grants No. 12174126 and No. 12104166), the Guangdong Basic and Applied Basic Research Foundation (Grant No. 2024B1515020018), the Science and Technology Program of Guangzhou (Grant No. 2024A04J3004), and the Open Fund of Key Laboratory of Atomic and Subatomic Structure and Quantum Control (Ministry of Education).

J.L.D. and E.W.L. contributed equally to this work.

\end{acknowledgments}

\normalem
\bibliography{reference}

\end{document}